\documentclass[aps,floatfix,nofootinbib,preprint]{revtex4}

\usepackage{graphicx}
\usepackage{epic}
\usepackage{eepic}
\usepackage{latexsym}
\usepackage{amssymb,amsmath}

\newcommand{\eq}[1]{(\ref{#1})}
\newcommand{\be}{\begin{equation}}
\newcommand{\ee}{\end{equation}}
\newcommand{\bea}{\begin{eqnarray}}
\newcommand{\eea}{\end{eqnarray}}

\newcommand{\vs}[1]{\vspace{#1 mm}}
\newcommand{\hs}[1]{\hspace{#1 mm}}

\def\a{\alpha}
\def\b{\beta}
\def\cc{\gamma}
\def\C{\Gamma}
\def\d{\delta}

\def\e{\epsilon}

\def\f{\phi}
\def\fr{\frac}

\def\L{\Lambda}
\def\m{\mu}
\def\n{\nu}

\def\r{\rho}
\def\s{\sigma}

\def\z{\zeta}
\def\O{\Omega}

\def\del{\partial}

\let\bm=\bibitem
\def\nn{\nonumber}

\begin{document}

\title{Big-Bang is a Boundary Condition} 

\author{Ali Kaya}

\email[]{alikaya@tamu.edu}
\affiliation{\vs{3}Department of Physics and Astronomy, Texas A\&M University, College Station, TX 77843, USA \vs{10}}

\begin{abstract}

\vs{5}

There is a common expectation that the big-bang singularity must be resolved in quantum gravity but it is not clear how this can be achieved. A major obstacle here is the difficulty of interpreting wave-functions in quantum gravity. The standard quantum mechanical framework  requires  a notion of time evolution and a  proper definition of an invariant inner product having a probability interpretation,  both of which are seemingly problematic in quantum gravity. We show that these two issues  can actually be solved by introducing the embedding coordinates as dynamical variables à la Isham and Kuchar. The extended theory is identical to general relativity but has a larger group of gauge symmetries. The Wheeler-DeWitt equations describe the change of the wave-function from one arbitrary spacelike slice to another, however the constraint algebra makes  this evolution purely kinematical and furthermore enforces the wave-function to be constrained in the subspace of zero-energy states. An inner product can also be introduced having all the necessary requirements. In this formalism big-bang appears as a finite field space boundary on which certain boundary conditions must be imposed for mathematical consistency. We explicitly illustrate this point both in the full theory and  in the minisuperspace approximation.

\end{abstract}

\maketitle

\section{Introduction}

The big-bang appears as a naked spacelike singularity in cosmological solutions of general relativity. As shown by various singularity theorems, its existence seems inevitable due to the cosmic expansion of the universe and the nature of gravity as an attractive force (see \cite{bgv} for a recent, simple but elegant example of a singularity theorem that shows how sufficiently fast expansion can yield  incompleteness in null and timelike past directions; see also \cite{yeni3} that discusses  the limitations of the results of \cite{bgv}). The classical theory looses its predictability near the big-bang singularity and one would naturally expect this issue to be resolved after quantization. Since we do not have a complete consistent theory describing quantum gravity, there are only heuristic arguments for how the big-bang singularity could be resolved. Here, based on the canonical quantization, we also would like to propose an intuitive (but still mainly heuristic)  picture of the big-bang singularity in quantum gravity. 

There are two well known proposals that deal with the big-bang singularity, or to be more precise with the beginning of the universe in quantum cosmology;  the no boundary  \cite{qc1} and the tunneling \cite{qc2} prescriptions. Both of these methods offer a way of calculating the initial wave-function of the universe using Euclidean methods; the first involves  a Euclidean gravitational path integral and the second uses the tunneling through a barrier via Euclidean time. However, it looks like a Lorentzian path integral calculation based on the Picard-Lefschetz theory gives a completely different picture \cite{qc3,qc4}. 

We think that these considerations involving the wave-function of the universe in quantum gravity, both the Euclidean and the Lorentzian approaches, have two major drawbacks. As is well known, the standard canonical quantization of general relativity gives the Wheeler-DeWitt (WDW) equation that has no explicit time derivative. This results the so called  problem of time in quantum gravity, i.e. it is not clear how to define a time evolution either at a fundamental level or maybe as an emergent phenomena. Related to this issue, it is also unclear how to introduce an invariant inner product having a viable probability interpretation. Note that the standard quantum mechanical treatment is based on a notion of time evolution and an invariant inner product. Most of the time, the physics extracted from the wave-function of the universe is based on a semi-classical/WKB reasoning where the wave-function takes the form $\Psi=A\exp(iS)$ and $S$ satisfies the classical Hamilton-Jacobi equation. It is then possible to define a conserved current but as discussed in \cite{wu} this does not yield a probability density  because the metric on the superspace is not positive definite. Furthermore, one would expect the semi-classical approximation to break down near the big-bang singularity. In any case, it is obvious that without a viable solution to {\it the problems of time and inner product,}  it is not possible to properly interpret the wave-function of the universe obtained in one way or another. 

There are quite a large number of different proposals tackling the problem of time in quantum gravity, see \cite{k1} for a relatively recent  extensive review. One approach is to use a field in the theory as an emergent time parameter that  keeps track of time evolution. This involves a gauge fixing and solving the Hamiltonian constraint for this variable's conjugate momentum, which  deparametrize the theory at the classical level. Then, the remaining degrees of freedom are quantized in the standard way, where the dynamics  is now governed by a non-vanishing Hamiltonian.  Not all field variables can play the role of time. For example, in a flat minisuperspace quantum cosmology of a self-interacting scalar field, the scale factor of the universe can be used as an emergent time  only if the scalar potential is positive definite \cite{ak1}. In  many cases, gauge fixing breaks down in some region in field space, as a result  the field ceases to be an honest time variable. To avoid this problem, one simply assumes an evolution outside that region and obviously this is not a satisfactory solution to the problem of time (as discussed  in \cite{yeni2},  the classical time may also emerge in the $\hbar\to0$ limit of the quantum probabilities). 
 
Naturally, one focuses on finding a time variable in dealing with the problem of time, however this conflicts with the principle of general covariance. Namely, choosing a time variable means specifying a particular Lorentzian foliation of the spacetime that explicitly breaks the general covariance.  Therefore, the problem of time must actually be reconsidered as {\it the problem of defining evolution from an arbitrary initial spacelike slice to an arbitrary final one.}  Needless to say, the canonical analysis requires a time variable (and hence a breakdown of the covariance) and it might be possible to change the originally chosen time coordinate, say,  by means of canonical transformations. Nevertheless  it is desirable to have a manifestly covariant approach that would presumably offer a much more clearer picture of general relativistic time evolution.

There is a framework developed by Isham and Kuchar in \cite{ik1,ik2} that enlarges the configuration space of a given theory by adding the embedding coordinates as dynamical variables (the embedding coordinates parametrize Lorentzian foliations in a spacetime). Their main aim was to simplify the algebra of constraints in general relativity that has "field dependent" structure constants. The analysis in \cite{ik2} was carried out directly in the Hamiltonian formalism but the diffeomorphism invariance was completely fixed from the beginning by choosing the Gaussian normal coordinates, which is an important disadvantage from the covariance point of view. In a recent work \cite{akn}, we have shown that one can actually start from a Lagrangian formulation by extending the configuration space  without the need of any prior gauge fixing  (see also \cite{ikg1,ikg2} that relax the necessity of imposing an initial gauge).  The extended theory has a larger gauge symmetry  associated with two different types of diffeomorphisms, however  it is fully equivalent to general relativity, which can be seen by partially fixing a gauge that eliminates  the embedding coordinates. 

As shown in \cite{akn}, and as we will briefly review below, the canonical analysis of the extended general relativity preserves the spacetime covariance. After quantization, the  WDW equation corresponding to a constraint becomes Schrödinger-like and describes the evolution of the  wave-function from one arbitrary spacelike slice to another. Moreover, it is possible to define an invariant inner product that has an obvious probability interpretation.  As a peculiarity of general relativity,  the constraint algebra yields a purely kinematical time evolution and enforces the wave-function  to be constrained in the subspace of zero-energy states. In this paper we would like to study how the big-bang singularity appears in this framework and as we will see our findings indicate a fundamentally different picture than the common lore.

\section{Extended General Relativity and Its Canonical Quantization}

Let us denote the configuration space of general relativity by ${\cal M}_g$,  the space of all Lorentzian metrics  on the spacetime manifold $M$ (one may assume that $M$ is diffeomorphic to $R^4$ to avoid any topological complications). Following \cite{ik1,ik2}, we extend the theory by adding the configuration space of all Lorentzian foliations, call  this ${\cal M}_X$. We imagine that the spacetime manifold $M$ has coordinates  $X^\m$  and the foliations are parametrized by $y^\a\equiv(\tau,y^i)$, where $\tau$ is an intrinsic time variable. Then,  ${\cal M}_X$ can be parametrized by the embedding maps $X^\m(y^\a)$. Of course, whether the embedding is Lorentzian depends on the metric and the embedding coordinates must obey the conditions  $g_{\m\n}X'^\m X'^\n<0$ and $\det \,( g_{\m\n}\del_\a X^\m \del_\b X^\n)<0$, where the prime denotes the $\tau$ derivative. Hence  ${\cal M}_X$ becomes an {\it open set} in the space of all embeddings  (since the conditions involve strict inequalities). Note that $X^\m(y^\a)$ can also be thought to parametrize the motion of a 3-brane, where $X^\m$ and $y^\a$ are target space and world-volume coordinates, respectively (sometimes we will  use this brane terminology). As a result,  the extended configuration space becomes a warped product ${\cal M}_g\ltimes {\cal M}_X$, where  ${\cal M}_X$ has a nontrivial fibration over ${\cal M}_g$.

One may now consider the following action for the extended configuration space variables
\be\label{pga} 
S=\fr12 \int d^4y \,\sqrt{-\cc}\, R(\cc)
\ee
where 
\be\label{cc}
\cc_{\a\b}=\del_\a X^\m \del_\b X^\n g_{\m\n},
\ee
$\cc=\det \cc_{\a\b}$ and $R(\cc)$ is the Ricci scalar of $\cc_{\a\b}$. Here, one should treat both $g_{\m\n}$ and $X^\m$ as dynamical variables depending on $y^\a$. The extended theory has a larger gauge symmetry compared to general relativity, i.e. the action \eq{pga} is invariant under both the world-volume and the target space diffeomorphisms.   These can be written as
\bea
&&\d X^\m=k^\a\del_\a X^\m,\nn\\
&&\d g_{\m\n}=k^\a \del_\a g_{\m\n}\label{wvd}
\eea
and 
\bea
&&\d X^\m=l^\m,\nn\\
&&\d g_{\m\n}=-\del_\a l^\r X_\m^\a  g_{\r\n}  -\del_\a l^\r X_\n^\a  g_{\r\m}  \label{tsd}
\eea
where $k^\a$ and $l^\m$ are world-volume and target space vectors (both of which must be thought as functions of $y^\a$). One can partially fix the gauge by imposing $X^\m=\d^\m_\a y^\a$ that eliminates the embedding coordinates. This yields the standard action of general relativity with the remaining gauge symmetry being the usual diffeomorphism invariance. Therefore, the extended theory is actually equivalent to general relativity. 

The conjugate momenta can be obtained by varying the action \eq{pga}; $P_\m=\d S/\d X'^\m$ and $P^{\m\n}=\d S/\d g'_{\m\n}$ where as above the prime denotes the $\tau$ derivative.  The canonical pairs are $(X^\m,P_\n)$ and $(g_{\m\n}, P^{\r\s})$, and {\it the spacetime covariance} is preserved in the canonical analysis (while the world-volume covariance is broken). The Legendre transformation  gives (up to surface terms that may arise in applying integration by parts) 
\be
\int d\tau d^3 y\left[ P^{\m\n}g'_{\m\n}+P_\m X'^\m-{\cal L} \right]=0,
\ee
where the Lagrangian density is ${\cal L}=\sqrt{-\cc}\, R(\cc)/2$. Therefore the Hamiltonian vanishes identically (we ignore  possible surface terms in this work), which should be compared to the Hamiltonian of general relativity that becomes a combination of constraints (in the language of Dirac, the first vanishes strongly and the second weakly).  

Since the Hamiltonian vanishes identically, the dynamics is solely governed by the constraints. As discussed in  \cite{akn}, there are  primary constraints, which directly follow from the definitions of conjugate momenta, and their algebra implies some secondary constraints. All these are first class and can be written after some simplifications as 
\bea
&&n_\n P^{\m\n}=0,\nn\\
&&\del_i X^\m P_\m+(\del_i g_{\m\n}) P^{\m\n}=0,\nn\\
&&n^\m P_\m + (\d_n g_{\m\n})\, P^{\m\n}=0,\label{rgrc}\\
&&\Phi_i\equiv D_j \left(\fr{1}{\sqrt{h}} P^j{}_i\right)=0,\nn\\
&&\Phi\equiv \fr{2}{\sqrt{h}}\left[P^{ij}P_{ij}-\fr12 P^2\right]-\fr12 \sqrt{h}R^{(3)}=0.\nn
\eea
Here, $n^\m$ is the future pointing  unit normal vector of the foliation that is uniquely determined by the equations $n_\m \del_i X^\m=0$ and $n^\m n^\n g_{\m\n}=-1$, where $\del_i=\del/\del y^i$. Also, 
\be
h_{ij}=\del_i X^\m \del_j X^\n g_{\m\n}, 
\ee
$h=\det(h_{ij})$, $D_i$ is the covariant derivative and $R^{(3)}$ is the Ricci scalar of $h_{ij}$, 
\bea
&&P_{ij}=\del_i X^\m \del_j X^\n P_{\m\n},\nn\\
&&P=P^{ij}h_{ij}.
\eea
We use $\d_n g_{\m\n}$ to denote  the change in the metric given in \eq{tsd} with $l^\m=n^\m$. In these equations $(i,j)$ indices are manipulated by the induced metric $h_{ij}$.  

\begin{figure}
	\centerline{\includegraphics[width=9cm]{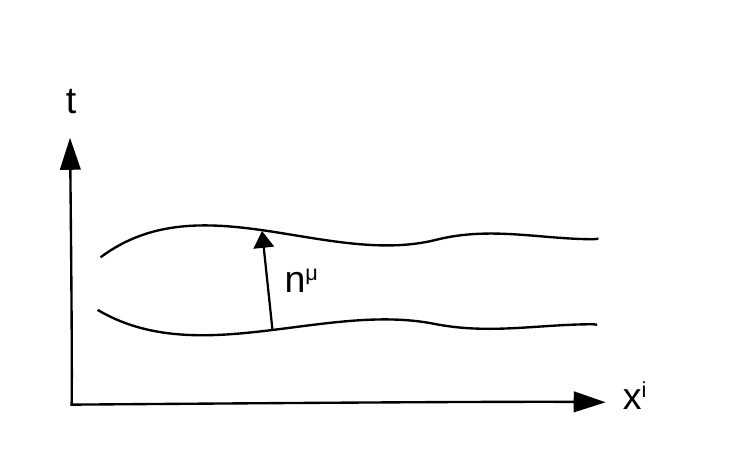}}
	\caption{An initial spacelike slice and its deformation along its normal vector.} 
	\label{fig1}
\end{figure}

The first equation in \eq{rgrc} shows that the hypersurface non-tangent metric components are non-dynamical. The second and the third equations generate transformations that are tangent and normal to a given spacelike surface, hence the second one describes time evolution, or to be more precise the deformation of an initial spacelike slice along its normal direction (for an illustration see Fig \ref{fig1}). As mentioned earlier, this is a desired description of time evolution in a general relativistic context, i.e. in this formulation one can start from an arbitrary initial spacelike surface and  consider the evolution to another slice. The fourth equation generates the change of the induced metric under a spatial coordinate transformation and the last equation is the new Hamiltonian constraint; below we will discuss its importance in quantum theory. 

Let us now work out the implications of  the constraints in the canonical quantization where one defines the conjugate momentum operators as $P_\m=-i\d/\d X^\m$ and $P^{\m\n}=-i\d/\d g_{\m\n}$. The wave-function can be taken as a functional $\Psi=\Psi[X^\m,g_{\m\n}]$, where the configuration variables are defined on a constant $\tau$-slice,  $X^\m=X^\m(y^i)$ and $g_{\m\n}=g_{\m\n}(y^i)$. The constraints are now implemented as operators in \eq{rgrc} annihilating $\Psi$  (placing the momenta to the rightmost  position seems to be a natural ordering for many of the following results to hold). It is useful to decompose the metric with respect to  the tangent and  non-tangent surface components, as a result one can write $\Psi=\Psi[X^\m,h_{ij},g_{\m\n}m^\n]$ where $m^\m$ is chosen so that $(\del_i X^\m,m^\n)$ forms a basis in the tangent space (specifically one has $n_\m m^\m\not=0$). The first constraint in \eq{rgrc} implies  that the wave-function must not depend on $g_{\m\n}m^\n$, i.e.  
\be\label{wfxh}
\Psi=\Psi[X^\m,h_{ij}]. 
\ee
The second and the fourth constrains ensure the invariance of $\Psi$ under a spatial coordinate change $y^i\to y^i+k^i$ both extrinsically $\d X^\m=k^i\del_i X^\m$, $\d h_{ij}=D_i k_j+D_j k_i$ and intrinsically $\d X^\m=0$, $\d h_{ij}=D_i k_j+D_j k_i$. Specifically, the second constraint can be shown to imply 
\be\label{sd}
\Psi[X^\m(y^i+k^i),h_{ij}+{\cal L}_k h_{ij}] -\Psi[X^\m(y^i),h_{ij}]=0
\ee 
to the first order in $k^i$.

The third constraint can be used to obtain the wave-function on a new spacelike slice given by the embedding $X^\m+\e(y^i)\,n^\m$, where $\e(y^i)$ is an arbitrary deformation parameter.   The third and the fifth constraints in \eq{rgrc} actually originate from the same primary constraint.  To be more precise, the primary constraint which follows from the  noninvertibility of the momentum-velocity map  is
\be
n^\m P_\m + (\d_n g_{\m\n})\, P^{\m\n}+\Phi=0. \label{ham}
\ee 
Obviously $\Phi$ can be identified as the Hamiltonian\footnote{Indeed, the analogue of \eq{ham} becomes the Hamiltonian in a  self-interacting scalar field theory \cite{akn}. } of the theory (note that $\Phi$ is quadratic in momenta). Remarkably in general relativity, the commutator of \eq{ham} with the first constraint in \eq{rgrc} implies $\Phi=0$. Consequently the condition $\Phi\Psi=0$ means $\Psi$ must belong to the subspace of {\it zero-energy states}. This is an intriguing result that fundamentally distinguishes general relativity from an ordinary field theory, which does not have the  first constraint in \eq{rgrc}. Note that $\Phi$ is a hyperbolic operator since the metric on the superspace is not positive definite. The time evolution of the wave-function \eq{wfxh} becomes {\it trivial} as the third constraint implies 
\be\label{tt}
 \Psi[X^\m+\e\, n^\m,h_{ij}]- \Psi[X^\m,h_{ij}]=0,
\ee
to first order in $\e(y^i)$. So to obtain the wave-functional on a new slice one only needs to deform the surface by keeping the induced metric constant. 

To complete the quantization, one should introduce an invariant inner product that has a viable probability interpretation. In \cite{akn}, we studied this problem in detail for a parametrized scalar field and observe that the quantum equivalence of the usual and the parametrized theories requires to define the inner product over the scalar field degrees of freedom where the embedding coordinates  are kept intact. This suggests 
\be\label{ip} 
\left<\Psi|\Omega\right>=\int D h_{ij} \, \Psi^*[X^\m,h_{ij}]\,\Omega[X^\m,h_{ij}] ... 
\ee
where the dotted terms represent  gauge fixing and Faddeev-Popov terms corresponding to the symmetry 
\be
h_{ij}\to h_{ij}+{\cal L}_k h_{ij}=h_{ij}+D_i k_j+D_j k_i,
\ee
where ${\cal L}_k h_{ij}=k^l\del_l h_{ij}+ h_{il}\del_j k^l + h_{jl}\del_i k^l$ is the Lie-derivative (our implementation here is similar to \cite{woodard}, more on this below). Here we treat $h_{ij}$ as the main dynamical variable and the path integral  is over all positive definite metrics that can be defined on the three-dimensional constant $\tau$-slice, we call this space $\Sigma_h$. This inner product, which is invariant under spatial coordinate changes and  time translations  as guaranteed by \eq{sd} and \eq{tt}, implies the obvious probability interpretation, i.e. $|\Psi[X^\m,h_{ij}]|^2$ defines an honest probability distribution over $\Sigma_h$. 

Note that in an improvised path integral quantization, one may naively carry out  a Lorentzian path integral over 4-dimensional metrics that are fixed on an initial and final slices in an attempt to obtain the propagator and the evolution of the wave-function. However, the constraints dictate that the states have zero energy and their time evolution is trivial, consequently it is enough to deal with a Euclidean path integral over the three dimensional metrics as in \eq{ip}. 

\section{The Big-Bang as a Boundary Condition}

In homogeneous and isotropic cosmological solutions of general relativity, the scale factor of the universe $a$ can be used to describe the time evolution and for expanding backgrounds the big-bang appears as a special past instant $a=0$. Most of the time this behavior leads to the idea that the universe has a special beginning. However, as we have discussed in the previous section, the time evolution in the canonical quantum gravity can be described with respect to the embedding coordinates (or to be more precise with respect to the corresponding spacelike slices) and there is no special instant about the time evolution dictated by \eq{tt} like big-bang. 

To discuss how the big-bang might appear in this framework, let us parametrize $h_{ij}$ as follows\footnote{This parametrization depends on the coordinates $y^i$ of the foliations. One can assume $y^i$ to be the standard Cartesian coordinates of $R^3$ since in this work we assume all manifolds are topologically trivial.} 
\be\label{accp}
h_{ij}=a^2 \cc_{ij},
\ee
where $\det(\cc_{ij})=1$, $\det(h_{ij})=a^6$ and $a>0$, and . Since $\cc_{ij}$ is symmetric and belongs to $SL(3)$,  the space of all positive definite metrics $\Sigma_h$ is the space of all $ R^+\times SL(3)/SO(3)$ fibers over the constant $\tau$-slice up to gauge transformations. From the structure of the standard cosmological solutions  such as  radiation or matter dominated universes, the big-bang can be associated with the collapse of a three dimensional constant time slice to zero size.  Hence a singular metric $h_{ij}$ which has the following property
\be\label{bbtm}
a(y^i)=0\hs{5} \exists\, y^i
\ee
(i.e. the scale factor $a(y^i)$ vanishes for some $y^i$) can be said to be the {\it big-bang type} (the standard  big-bang of the  radiation or matter dominated universes correspond to  $a(y^i)=0$  $\forall$ $y^i$). Such metrics constitute the {\it finite boundary} of the space $R^+\times SL(3)/SO(3)$, i.e. $\{0\}\times SL(3)/SO(3)$ (there is also a boundary at infinity since both $R^+$ and $SL(3)$ are non-compact). The  decomposition \eq{accp} is preserved under diffeomorphisms since the fields transform as 
\bea
&& \d a = k^i\del_i a+ \fr13  (\del_i k^i) a\nn\\
&&\d \cc_{ij}={\cal L}_k \cc_{ij}-\fr23 (\del_l k^l)\cc_{ij}, \label{cort}
\eea
i.e. $a$ and $\cc_{ij}$ do not mix with each other. 

The wave-function obeys the functional differential equation on $\Sigma_h$
\be
\Phi\Psi=0
\ee
and therefore the theory {\it requires} some boundary conditions to be imposed at the finite field space boundary (note that in principle changing the boundary conditions actually redefines the theory). To understand the structure of the field space finite boundary better, one can introduce the conjugate pairs $(a,P_a)$ and $(\cc_{ij},\s^{kl})$ by decomposing $P^{ij}$ as
\bea
&&P^{ij}=\fr13 h^{ij} P + \fr{1}{a^2}\s^{ij},\nn\\
&&P_a=\fr{2}{a} P,
\eea
where $\cc_{ij}\s^{ij}=0$. These obey the canonical brackets 
\bea
&& \{a(y_1),P_a(y_2)\}=\d^3(y_1-y_2),\nn\\
&&\{\cc_{ij}(y_1),\s^{kl}(y_2)\}=\left[\fr12\left(\d^k_i\d^l_j+\d^l_i\d^k_j\right)-\fr13\cc_{ij}\cc^{kl}\right]\d^3(y_1-y_2).
\eea
In the new variables the constraint $\Phi$ becomes 
\be\label{phia}
\Phi=-\fr{1}{12 a} P_a^2+\fr{2}{a^3}\cc_{ij}\cc_{kl}\s^{ik}\s^{jl}-\fr12 \sqrt{h}R^{(3)}.
\ee
This defines a metric in the configuration space\footnote{Note that given a nonlinear sigma model Lagrangian $L=\fr12 G_{AB}\dot{x}^A \dot{x}^B$ that has the metric $G_{AB}$, the corresponding Hamiltonian becomes $H=\fr12 G^{AB}p_A p_B$.} (the superspace) which can be read as 
\be\label{ssmet}
dS^2[a,\cc_{ij}]=\int d^3 y \left[-12 a (\d a)^2+\fr12 a^3 \cc^{ij}\cc^{kl}(\d \cc_{ik} \d \cc_{jl})\right],
\ee
where $(a,\cc_{ij})$ is a point and $(\d a, \d\cc_{ij})$ is a cotangent space one-form in the superspace. Note that  \eq{ssmet}  is a direct product of a "radial" direction with the left invariant $SL(3)$ metric on the coset $SL(3)/SO(3)$.  The metric \eq{ssmet} shows that the field space boundary defined by \eq{bbtm} is genuinely at a finite distance and for instance  it  cannot be avoided  by a field redefinition like $a=e^\z$ that seemingly pushes it to infinity $\z\to-\infty$. 

In addition to imposing boundary conditions at \eq{bbtm} one should also work out the path integral measure in \eq{ip} that defines the inner product.  This generally requires introducing a (positive definite) metric in the field space. In our case, one has already the superspace metric \eq{ssmet} but it is not positive definite.  One may for instance introduce the standard metric
\be
dS^2[h_{ij}]=\int d^3 y\, \sqrt{h} \,h^{ij}h^{kl} (\d h_{ik} \d h_{jl})
\ee
or its generalization that involves other index contractions, however there is no canonical choice here.  In any case, if one demands the field space metric to be covariant and ultralocal, and use the parametrization \eq{accp}, the path integral measure can be written in general as 
\be\label{mesacc}
Dh_{ij}=\prod_{y^i} w(a(y^i))\,da(y^i)\, d\cc(y^i)
\ee
where $w(a)$ is a weight function depending on $a$ and $d\cc$ represents the (5-dimensional) left invariant Haar measure on $SL(3)/SO(3)$, i.e.
\be\label{volform}
d\cc=\mathrm{Tr} \left(L\wedge L\wedge L\wedge L\wedge L\right),
\ee
where  $L^{i}{}_j=\cc^{ik}d\cc_{kj}$ denotes the left invariant one-forms.\footnote{Here one can imagine $\cc_{ij}=\cc_{ij}(\a_A)$ for some coordinates $\a_A$, $A=1,..,5$, parametrizing the coset $SL(3)/SO(3)$ like  $\cc_{ij}=\exp(\a_A T^A_{ij})$, where $\mathrm{Tr}\, T^A_{ij}=0$ and $T^A_{ij}=T^A_{(ij)}$.} 

At this point, it is crucial to ensure that the measure \eq{mesacc} is invariant under  coordinate transformations \eq{cort}. One can show that the left invariant one-form $L^i{}_j$ transforms like a tensor $\d L^i{}_j={\cal L}_k L^i{}_j$  (note that $d\cc_{ij}$ is a one-form in the coset space $SL(3)/SO(3)$ not in the spatial constant $\tau$-slice) and consequently the volume form \eq{volform} transforms like a scalar
\be
\d (d \cc) =k^i\del_i (d\cc).
\ee
The Jacobian of this change in the path integral measure \eq{mesacc} is a field independent determinant of the derivative operator. Similarly, if one assumes that the weight function has a simple power law dependence
\be\label{wan}
w=a^n
\ee
the path integral measure for the scale factor becomes either $\prod_{y^i} da^{n+1}(y^i)$ ($n\not=-1$) or $\prod_{y^i} d\ln a(y^i)$ ($n=-1$). From the following behavior under an infinitesimal coordinate change
\bea
&& \d a^{n+1}=k^i\del_i a^{n+1}+\fr{n+1}{3}(\del_ik^i) a^{n+1},\nn\\
&&\d \ln a = k^i\del_i \ln a +\fr13 \del_ik^i,
\eea
again the measure can be seen to pickup a trivial Jacobian. Therefore, the path integral measure \eq{mesacc} is invariant under spatial diffeomorphisms if the weight function has the form  \eq{wan}. As discussed in \cite{yeni1}, different choices of the weight functions can be related to each other by field redefinitions, which also correspond to different ordering prescriptions. Thus, in the calculation of certain physical quantities such as transition amplitudes all choices yield the same answer. However, as pointed out in \cite{mal},  the field redefinitions change the answer in the calculation of (cosmologically  relevant) correlation functions.

The final step in this construction is to eliminate the redundant gauge degrees of freedom in the path integral  \eq{ip} and one can apply the standard  Faddeev-Popov method to achieve this. The gauge fixing can be done by imposing, e.g. $G_j\equiv \del_i\cc_{ij}=0$ and the gauge fixed  inner product can be written explicitly as
\be\label{ipgf}
\left<\Psi|\Omega\right>=\int  \prod_{y^i} a(y^i)\,da(y^i)\, d\cc(y^i)\, \d(G_j)\, \det M_{ij}\, \Psi^*[X^\m,h_{ij}]\,\Omega[X^\m,h_{ij}] 
\ee
where $M_{ij}=\d_k G_j/\d k^i$, $\d_k G_j$ is the change in $G_j$ under a diffeomorphism generated by $k^i$ and  $ \det M_{ij}$ is the  Faddeev-Popov determinant, which can be evaluated by introducing the ghost fields. 

One may be puzzled with the integral over the scale factor $a$ in \eq{ipgf} since one would expect to see only two real degrees of freedom for gravity, which is carried out by the trace-free and transverse (after gauge fixing) variable $\cc_{ij}$. However, the states now obey the condition $\Phi\Psi=0$ hence counting of the physical degrees of freedom agrees with a spin-2 field  (e.g. the constraint $\Phi=0$ can be used to eliminate $a$ in the classical theory). 

Obviously, one should now try to constrain the theory more to determine the weight function \eq{wan} and possible  boundary conditions at the the big-bang type metrics \eq{bbtm} (surface terms at infinity can usually be discarded by assuming suitable fall-off conditions). Of course one may prefer not to impose any boundary conditions but this does not change the fact that there is a finite field space boundary for a functional differential equation and hence the theory is not uniquely defined unless one imposes certain boundary conditions.\footnote{Since $\d/\d a$ is a  timelike tangent vector with respect to the superspace metric \eq{ssmet}, the boundary \eq{bbtm} can be viewed as a spacelike surface in the field space. We prefer to use the term boundary conditions instead of  initial conditions since the latter can be confused with the spacetime evolution.} Since the constraint equation  $\Phi\Psi=0$ involves the second order derivative $\d^2/\d a^2$, one can for instance impose the standard Dirichlet or Neumann conditions
\bea
&&\Psi_D[X^\m(y^i),a(y^i),\cc_{ij}(y^i)]=0\hs{5} \mathrm{when}\hs{5} \left\{ a(y^i)=0\hs{5}\exists\, y^i \right\},\nn\\
&&\fr{\d}{\d a(y^i)}\Psi_N[X^\m(y^i),a(y^i),\cc_{ij}(y^i)]=0\hs{5} \mathrm{when}\hs{5}\left\{ a(y^i)=0\hs{5}\exists\, y^i\right\}, \label{dn} 
\eea
or consider a more general mixed boundary conditions
\be\label{mixedbc}
\a\, \Psi[X^\m(y^i),a(y^i),\cc_{ij}(y^i)]+\b \,\fr{\d}{\d a(y^i)}\Psi[X^\m(y^i),a(y^i),\cc_{ij}(y^i)]=0\hs{2} \mathrm{when}\hs{2}\left\{ a(y^i)=0\hs{2}\exists\, y^i\right\}, 
\ee
where $\a(y^i)$ and $\b(y^i)$ are arbitrary functions. 

One may also want some of the operators in the theory to be Hermitian. This is especially necessary for an operator corresponding to a physical observable (although determining physical observables in quantum gravity is quite challenging). For instance,  to have $P_a^\dagger=P_a$ one should choose $w=1$ and apply the Dirichlet boundary condition in \eq{dn}. Note that the gauge fixing term and the corresponding Faddeev-Popov determinant in \eq{ipgf} do not depend on $a$. On the other hand, one may demand $\Phi^\dagger=\Phi$; this condition would be necessary for the invariance of the inner product if the time evolution would have been dictated by \eq{ham} (apparently, this is not  required in general relativity but one may still want to impose it since the  solutions $\Phi\Psi=0$ depend on the properties of the operator $\Phi$ under the inner product \eq{ipgf}). To satisfy  $\Phi^\dagger=\Phi$, one should set $w=a$ and impose one of the boundary conditions in \eq{dn} or \eq{mixedbc}. Another interesting potential observable is the trace of the extrinsic curvature $K=-2P/a^3=-P_a/a^2$, which generalizes the Hubble parameter of a homogeneous and isotropic background to a general universe.  To have $K^\dagger=K$ one should choose $w=a^2$ and impose the Dirichlet boundary condition in \eq{dn}. Note that each option mentioned above actually yields a different theory. 

It is instructive to compare the above construction with the seminal work of DeWitt \cite{dewitt},  where two crucial differences stand out. First, in \cite{dewitt} DeWitt implemented the Klein-Gordon inner product, which is not positive definite.  We think that using the the Klein-Gordon inner product inevitably leads to the so called third quantization where the wave-function $\Psi$ becomes an operator and the positive/negative norm modes become the expansion coefficients of the creation/annihilation operators (recall how one reaches out the field theory interpretation for a quantum mechanical wave-function obeying the Klein-Gordon equation). The second difference is related to implementation of the boundary conditions where in \cite{dewitt} DeWitt proposed to impose $\Psi=0$ on singular three-geometries, which corresponds to the Dirichlet boundary condition in \eq{dn}. Here, we see that mathematical consistency actually allows more possible choices and as we discuss  in a minisuperspace model below in some cases it might be necessary to use this freedom to make sense of the theory. One may think that imposing $\Psi=0$ is physically viable since this would imply the probability of finding a singular geometry is zero. However, $|\Psi|^2$ defines a {\it probability density} not a probability. As we pointed out above,  by the parametrization \eq{accp} singular metrics form the {\it boundary} of the configuration space, hence simply by the measure of the spaces (bulk vs. boundary) it is not mathematically meaningful\footnote{To explain this point in an elementary example, imagine the quantum mechanical problem of a free particle in an infinite well $0<x<L$. Just by the measure, it is meaningless to ask about the probability of finding the particle at $x=0$. The wave-function only gives the probability of finding the particle in a {\it finite} interval.} to talk about the probability of finding a singular metric (or a set of singular metrics).  

Let us now finally see the implications of our findings in a simple minisuperspace approximation. For a real self-interacting scalar field $\f$ on a cosmological spacetime
\be
ds^2=-N^2d t ^2+a^2d\vec{x}^2\label{cs}
\ee
the reparametrized minisuperspace action corresponding to \eq{pga} can be written as \cite{akn} 
\be\label{ms}
S=\int\, d\tau\, a^3\left[-\fr{3}{Nt'}\fr{a'^2}{a^2}+\fr{1}{2Nt'}\f'^2-Nt'V\right],
\ee
where the prime denotes $\tau$-derivative and $V$ is the scalar potential. Here, $t$ is the embedding coordinate and one treats $N$, $t$, $a$ and $\f$ as the independent degrees of freedom depending on $\tau$. One can see by imposing the gauge $t=\tau$ that \eq{ms}  gives the usual minisuperspace action showing the equivalence of the two theories. 

The primary constraints that follow from the action \eq{ms} become
\bea
&&P_N=0,\nn\\
&&P_t + N {\cal H}=0,\label{mspc}
\eea
where 
\be\label{nf}
{\cal H}=-\fr{1}{12a}P_a^2+\fr{1}{2a^3}P_\f^2+a^3V. 
\ee
The first constraint shows that the wave-function does not depend on $N$ and thus takes the form  $\Psi(a,\f;t)$, and the second constraint gives the Schr\"{o}dinger equation with respect to the embedding time
\be\label{schr}
i\frac{\del}{\del t} \Psi = N {\cal H} \Psi. 
\ee
As mentioned above, the inner product must be introduced over the variables other than the embedding coordinate, and thus one should  define
\be\label{ip2} 
\left<\Psi|\O\right>=\int_{-\infty}^{+\infty} \int_0^\infty w(a)\, da\, d\f \, \Psi^*\O
\ee
where $w(a)$ is a possible wight function (the existence of a weight function in the inner product has also been considered in \cite{yeni1},  where this freedom is fixed by imposing the Hamiltonian to be Hermitian, which is similar to our treatment below). The inner product becomes invariant under time translations $t\to t +\e$ assuming that ${\cal H}$ is a Hermitian operator ${\cal H}^\dagger ={\cal H}$. In that case,  $|\Psi|^2\, w(a)\, da \,d\f$ gives a probability distribution over $a$ and $\f$.  

Obviously, the big-bang can be identified with $a=0$ at which certain boundary conditions must be imposed, especially to have ${\cal H}^\dagger ={\cal H}$ (one should also assume suitable falloff conditions as $a\to + \infty$ and $\f\to\pm\infty$, which can treated by using test functions as in \cite{dewitt};  the generalization of this treatment for an arbitrary measure has been provided in \cite{yeni1}). Similar to \eq{dn} and \eq{mixedbc}, one can impose Dirichlet,  Neumann or mixed boundary conditions 
\bea
&&\Psi_D(0,\f)=0,\nn\\
&&\fr{\del}{\del a} \Psi_N(0,\f)=0,\label{mbc} \\
&& \a \Psi(0,\f) + \b \fr{\del}{\del a} \Psi(0,\f)=0,\nn
\eea
which ensures\footnote{As the minisuperspace model is closely connected to an ordinary quantum mechanical system, we prefer to impose  ${\cal H}^\dagger ={\cal H}$, however, as discussed above, other choices are also possible giving different quantizations of the same system.} ${\cal H}^\dagger ={\cal H}$ provided one also chooses $w(a)=a$. Note that when ${\cal H}^\dagger ={\cal H}$ there always exists  a self-adjoint extension since ${\cal H}$ is real. 

So far this is an ordinary quantum mechanical system having the  Schr\"{o}dinger equation \eq{schr} (it is possible to set $N=1$  by introducing the proper time parameter $dt_P=Ndt$) and the inner product \eq{ip2}, but the constraint algebra demands more. The commutator of the two terms in \eq{mspc}  requires
\be
{\cal H}\Psi=0\label{wdwo}
\ee
and thus the states are confined in the zero-energy subspace and their time evolution is trivial (nevertheless, the standard quantum mechanical rules  of calculating probabilities or expectation values of observables still apply). 

To illustrate the importance of different possible boundary conditions, let us further consider a vacuum spacetime with a cosmological constant $\L$ so that \eq{nf} becomes 
\be\label{nf2}
{\cal H}=-\fr{1}{12a}P_a^2+a^3\L. 
\ee
The wave-function obeys (as noted above, we place the momenta to the right most position in operator ordering)
\be\label{schr1}
\fr{d^2 \Psi}{da^2}+12 \L a^4 \Psi=0, 
\ee
and it must be normalized so that
\be\label{norm} 
\left<\Psi|\Psi\right>=\int_0^\infty  | \Psi|^2 \, a\, da=1.
\ee
The two solutions of \eq{schr1} can be found in terms of the Bessel functions. One can see that for $\L\geq0$, the states are not normalizable due to their oscillating behavior as $a\to\infty$, therefore the minisuperspace approximation is not well defined. For $\L<0$, one solution is exponentially decaying and the other one is growing, and the later is not normalizable giving the unique wave-function
\be\label{ex1}
\Psi=C\, \sqrt{a}\, K_{1/6}\left(\fr{\sqrt{12|\L|}}{3} a^3\right),
\ee
where $C$ is a normalization constant.\footnote{Note that $K_{1/6}(x)\sim 1/x^{1/6}$ as $x\to0$, thus \eq{norm} also converges as $a\to0$ thanks to the presence of the weight function.} This is a rather trivial example of a quantum mechanical system having a one-dimensional  Hilbert space, still one can see that \eq{ex1} obeys the mixed boundary condition 
\be
\fr{d\Psi}{da}(0)=\fr{\C(-1/6)}{\C(1/6)}\left(\fr{\sqrt{12|\L|}}{3}\right)^{1/3}\Psi(0)
\ee
rather than the Dirichlet one. Note that at the classical level the vacuum Einstein equations for a cosmological ansatz are inconsistent for $\L<0$ since the Friedmann equation requires a positive energy density ($\L=0$ and $\L>0$ yield the flat and the de Sitter spaces). It turns out in the minisuperspace approximation it is only possible to make sense of the quantum mechanical description of the  $\L<0$ case.\footnote{This situation is very similar to the quantum mechanical description of a  {\it zero-energy} particle $E=0$ moving in the semi-infinite line $0\leq x<\infty$ in a constant potential $V_0$. In this system,  for $V_0>0$ (that gives $E<V_0$) one finds a tunneling wave-function that is exponentially decaying and for $V_0\leq0$ (that gives $E\geq V_0$) there is no properly normalizable state. The gravitational case has the opposite behavior since $P_a$ has the wrong sign in the Hamiltonian.}

\section{Conclusions} 

In this paper we show that in a proper quantum mechanical analysis where an appropriate time evolution and an invariant inner product yielding a viable probability interpretation are defined, the big-bang appears as a finite boundary in the configuration space of the three dimensional metrics at which certain boundary conditions must be imposed. It is important to underline that this assessment is different from the no boundary or tunneling proposals, which suggest ad hoc  ways of determining the {\it initial} wave-function of the universe. 

Is there a way to uniquely determine the boundary conditions?  One possible requirement is to impose certain operators to be Hermitian, and another  is to achieve the normalizability of the states. In the rather trivial example of an empty spacetime with a negative cosmological constant in the minisuperspace approximation, we see that the unique normalized wave-function obeys a mixed boundary condition. One may hope that, at least in the minisuperspace approximation, the boundary conditions can be  fixed by the mathematical consistency of the quantum mechanical description like the normalizability of the wave-functions. 

In any case, even if a suitable boundary condition is imposed, the wave-function of the universe can be an {\it arbitrary} vector; there is no {\it special initial state} to be  determined. In previous work \cite{ak1,ak2}, we have  studied the fate of the big-bang singularity in a {\it gauge fixed} minisuperspace model where the scale factor $a$ can be chosen as time and the big-bang appears as the first moment $a=0$. In that model the Hamiltonian degenerates to the zero operator as $a\to0$ and the  initial wave-function of the universe can be an {\it arbitrary} vector in the Hilbert space. Our findings here, based on a more general framework, are consistent with those results, i.e. in quantum gravity, at least in the canonical framework,  there is no special emphasis on the initial state of the universe. It looks like new physics is needed beyond  quantum mechanics and/or general relativity to understand the initial state of the universe unambiguously.

\end{document}